\documentclass[a4paper,fleqn,usenatbib]{mnras}

\usepackage{newtxtext,newtxmath}
\usepackage{CJKutf8}
\usepackage[T1]{fontenc}
\usepackage{ae,aecompl}

\usepackage{graphicx}	% Including figure files
\usepackage{amsmath}	% Advanced maths commands
\usepackage{amssymb}	% Extra maths symbols
\usepackage{multirow}

\title[Parental clusters and field planetary systems]{The signatures of the parental cluster on field planetary systems}

\author[M. X. Cai et al.]{
Maxwell Xu Cai (\begin{CJK*}{UTF8}{gbsn}蔡栩\end{CJK*}),$^{1}$\thanks{E-mail: cai@strw.leidenuniv.nl (MXC)}
Simon Portegies Zwart,$^{1}$ and Arjen van Elteren $^{1}$
\\
$^{1}$Leiden Observatory, Leiden University, PO Box 9513, 2300 RA, Leiden, The Netherlands\\
} 

\date{Accepted 2017 November 19. Received 2017 November 3; in original form 2017 September 17}

\pubyear{2017}

\begin{document}
\label{firstpage}
\pagerange{\pageref{firstpage}--\pageref{lastpage}}
 \maketitle

% Abstract of the paper
\begin{abstract}
Due to the high stellar densities in young clusters, planetary systems formed in these environments are likely to have experienced perturbations from encounters with other stars. We carry out direct $N$-body simulations of multi-planet systems in star clusters to study the combined effects of stellar encounters and internal planetary dynamics. These planetary systems eventually become part of the Galactic field population the parental cluster dissolves, which is where most presently-known exoplanets are observed. We show that perturbations induced by stellar encounters lead to distinct signatures in the field planetary systems, most prominently, the excited orbital inclinations and eccentricities. Planetary systems that form within the cluster's half-mass radius are more prone to such perturbations. The orbital elements are most strongly excited in the outermost orbit, but the effect propagates to the entire planetary system through secular evolution. Planet ejections may occur long after a stellar encounter. The surviving planets in these reduced systems tend to have, on average, higher inclinations and larger eccentricities compared to systems that were perturbed less strongly. As soon as the parental star cluster dissolves, external perturbations stop affecting the escaped planetary systems, and further evolution proceeds on a relaxation time scale. The outer regions of these ejected planetary systems tend to relax so slowly that their state carries the memory of their last strong encounter in the star cluster. Regardless of the stellar density, we observe a robust anticorrelation between multiplicity and mean inclination/eccentricity. We speculate that the ``Kepler dichotomy'' observed in field planetary systems is a natural consequence of their early evolution in the parental cluster.
\end{abstract}

% Select between one and six entries from the list of approved keywords.
% Don't make up new ones.

%%%%%%%%%%%%%%%%%%%%%%%%%%%%%%%%%%%%%%%%%%%%%%%%%%

%%%%%%%%%%%%%%%%% BODY OF PAPER %%%%%%%%%%%%%%%%%%

\section{Introduction}
The majority of stars, maybe even all, are born in a clustered environment \citep[e.g.,][]{2003ARA&A..41...57L,2003AJ....126.1916P,2012MNRAS.426L..11G}. Planets, so far as we know, are born orbiting stars. A planetary system is affected by this environment through encounters with other stars \citep{2009ApJ...697..458S, 2013MNRAS.433..867H, 2015MNRAS.448..344L,2016ApJ...816...59S,2017MNRAS.470.4337C}. To what degree can we recognize such early dynamical evolution in the orbital distribution of planets in a system at a later time depends on the magnitude of the perturbation and the exposure time \citep{2015MNRAS.451..144P}.

When a planetary system is perturbed externally by a passing star in the parental cluster, the perturbation may propagate to the inner system via secular evolution, and be preserved by the communal dynamical evolution of the planets in the system \citep{2017MNRAS.470.4337C}. If the planetary system becomes dynamically unstable due to such external perturbations, for example when an outer planetary orbit starts to cross a more inner orbit, this may result in the ejection of a planet. The reduced planetary system may still show signatures of this occurrence in the orbital elements of the surviving planets, such as in the inclination or the eccentricity. As a consequence reduced planetary systems may have different characteristics when compared to non-reduced planetary systems with happened to be born with the same number of planets.  For convenience, we make the distinction between \emph{rich planetary systems} (for systems with 3 or more planets) and \emph{poor planetary systems} (with fewer than 3 planets. This distinction is practical because rich planetary systems are by definition untouched/less affected, whereas poor planetary systems can either be poor from birth or reduced rich systems. From a theoretical perspective, there is no reason why planetary systems could not be born very rich (with $\gg 3$ planets) that remained rich even after reduction to $> 3$, but we ignore those cases here.

   We assume that planets form with small eccentricity and co-planar, much in the same way we have been perceiving planet formation since mid 18th-century \citep[e.g.,][]{1755anth.book.....K, de1809exposition, 1993ARA&A..31..129L}.  High relative inclinations are then hard to explain by their own internal dynamical evolution. Large deviations from circular orbits, however, can be explained by planet-planet interactions and possible ejections. Such ejections naturally lead to a reduction in the number of planets. In contrast, a wide variety of inclinations, either originating from an external source or by internal evolution, are not likely to result in the reduction of the system.
%when they dissolve, the member stars become part of the field population. Moreover, the planet formation process typically takes place within 3-10~Myr \citep[e.g.,][]{2011ARA&A..49...67W}, which is short compared with the typical dissolution timescales of star clusters \citep[e.g.,][]{1961AnAp...24..369H, 2003MNRAS.340..227B, 2011MNRAS.413.2509G, 2016MNRAS.455..596C}. As such, it is likely that planet formation occurs in star clusters. Based on our understanding that planetary systems probably have been perturbed in their birth clustered environment, and that these perturbations leave a signature that is preserved over a long time, we may be able to recognize the various characteristics in the observed population of planetary systems.

NASA's \emph{Kepler} mission observed an excess in the number of single-transit systems \citep[e.g.,][]{2011ApJS..197....8L,2012ApJ...758...39J}. This phenomenon was named the ``Kepler dichotomy'', because such an excess of single-transient systems is inconsistent with current theories of planet formation \citep[e.g.,][]{2013ApJ...775...53H}. Interestingly, these single-transit systems are often found to have large obliquities, whereas rich planetary systems tend to have a low average inclination, i.e., they tend to be coplanar \citep{2014ApJ...796...47M}. 
A simple solution would be that single-planet systems form differently than multi-planet systems, but
\cite{2015ApJ...807...44P,2015ApJ...806L..26V} argue that they are the leftovers of richer systems. They notice that rich systems discovered by Kepler tend to be fully packed and therefore only marginally stable. The ejection of several planets then naturally result in a poor-planetary system with a high obliquity. Similarly, \cite{2017MNRAS.468.3000M} argue that the inner part of a planetary system can be affected by either a giant planet in the outer part or a wide binary stellar companion. Planet-planet scattering and/or Lidov-Kozai mechanism \citep[e.g.,][]{1962AJ.....67..591K,1962P&SS....9..719L,1979A&A....77..145M} will excite the inclinations and eccentricities of the inner system, and eventually reduce the number of planets in the system. They also point out, however,  that dynamical evolution due to the outer planets alone is insufficient to explain the excess of \emph{Kepler}'s single-transit systems. \cite{2017MNRAS.469..171R} also suggest that single-transit systems may have inclined/non-transiting outer planets. On the other hand, a recent study by \cite{2017MNRAS.470.1750I} suggests that when the gas disk is present, planetary embryos grow and migrate inward. A resonance chain is formed when they capture each other into resonance. However, when the gas disk dissipates, the chain may break due to dynamical instability. Some planets may be induced to high inclinations, making then out of the line-of-sight and become undetectable for transit surveys. In this sense, single-transit systems may not be really single; rather, they may have hidden non-transiting companions.

In this study, we model the dynamical coevolution of planetary systems with their parental clusters using direct $N$-body simulations. We want to establish how stellar encounters affect the orbital inclinations and eccentricities of planetary systems when they are still in the parental cluster. Our results show that planetary systems formed within the half-mass radii of their parental clusters are much more likely to suffer from intensive external perturbations. Consequently, these systems are dynamically hotter, i.e., with more eccentric orbits and higher mutual inclinations. Additionally, these systems statistically have fewer planets compared to the systems form outside the half-mass radius of the cluster, in the sense that dynamical instabilities, initially triggered by intensive stellar encounters, leads to planet ejections. Based on these results, we argue that the ``Kepler dichotomy'' may be a signature of the dynamical evolution of the planetary systems in a young star cluster, and that the single observed planet is often accompanied by others in an inclined orbit (especially if this single-transiting planet has moderate-to-high orbital eccentricities, which indicates that the planetary system is dynamically hot). In contrast, rich planetary systems tend to exhibit much higher degrees of coplanarity, because they form in the outskirts of the parental clusters, where stellar densities are low and external perturbations are infrequent. We argue that both aspects are a natural consequence of the external perturbations these planetary systems endure, and their subsequent internal secular dynamical evolution.  These processes leave rich planetary systems planar and poor planetary systems with high inclinations.

This paper is organized as follows: the details of numerical modeling and initial conditions are presented in Section~\ref{sec:ic}; the results are shown in Section~\ref{sec:results}, followed by the discussions in Section~\ref{sec:discussion}. Finally, the conclusions are summarized in Section~\ref{sec:conclusions}.

%In this study, we propose that the inclinations are primordial, induced by stellar flybys when planetary systems are still in their parental clusters. Stars form in star clusters and stellar associations \citep[see, e.g., ][]{2003ARA&A..41...57L, 2003AJ....126.1916P}, and when they dissolve, the member stars become part of the field population. Moreover, the planet formation process typically takes place within 3-10~Myr \citep[e.g.,][]{2011ARA&A..49...67W}, which is short compared with the typical dissolution timescales of star clusters \citep[e.g.,][]{1961AnAp...24..369H, 2003MNRAS.340..227B, 2011MNRAS.413.2509G, 2016MNRAS.455..596C}. As such, it is likely that planet formation occurs in star clusters. Most presently known planets are found in the field, but their history in the parental clusters may leave signatures in the orbits of these planets \citep{2009ApJ...697..458S, 2013MNRAS.433..867H, 2015MNRAS.448..344L, 2016ApJ...816...59S, 2017arXiv170603789C}. In this paper, we focus on how external perturbations, mostly hyperbolic and nearly parabolic encounters, imprint in the subsequent dynamical process of field planetary systems. This paper is organized as follows: the details of numerical modeling and initial conditions are presented in Section~\ref{sec:ic}; the results are shown in Section~\ref{sec:results}, followed by the discussions in Section~\ref{sec:discussion}. Finally, the conclusions are summarized in Section~\ref{sec:conclusions}.

\section{Modeling and Initial Conditions}
\label{sec:ic}
We perform direct $N$-body simulations of star clusters including multi-planet systems.
Our objective is to study the departures of multi-planet systems from their initial configuration due to the combined effects of stellar perturbations and planet-planet scattering.
Planets in our simulations are assumed to be formed on a plane and with circular orbits. However, in time these canonical initial conditions deviate due to the combined effects of stellar encounters and planet-planet scatterings. The initial conditions for the host star cluster are sampled from \cite{1911MNRAS..71..460P} spheres with $N=2000$, $N=8000$, and $N=32,000$ stars, respectively, and the virial radius of all clusters are set to $R_{\rm vir} = 1$~pc. Stellar masses are randomly sampled from a broken power-law \cite{2001MNRAS.322..231K} with a minimum of $0.08 M_{\odot}$ and a maximum of $25 M_{\odot}$. This IMF results in a mean stellar mass of $\sim 0.55 M_{\odot}$.

The initial condition of planetary systems based on the ``EMS'' systems described in \cite{2007ApJ...666..423Z}, in which all planets in a planetary system is {\bf e}qual {\bf m}ass, {\bf s}eparated equally in terms of mutual Hill radii, hence the name. Since planetary systems are generally chaotic, we need to obtain our results statistically. Therefore, we initialize an ensemble of 200 identical EMS systems and assign them to solar-type stars ($M_{\rm star} = 1 \pm 0.02 M_{\odot}$) in the host cluster. The stellar density in the vicinity of a planetary system varies, depending on its location at the cluster. We adopt two EMS models (\textsc{Model J} and \textsc{Model E}). In \textsc{Model J} each planet has a mass equal to Jupiter ($\sim 10^{-3} M_{\odot}$); in \textsc{Model E} the planet mass is reduced by a factor of 1000, comparable to $\sim 1/3$ Earth mass ($\sim 10^{-6} M_{\odot}$). The semi-major axis of the innermost planet is set to $a_0 = 5.2$~AU, similar to Jupiter's orbit in the Solar system. Planets are placed on circular and coplanar orbits with semi-major axes $a = [5.2, 13.04, 32.7, 82.2, 206.2]$~AU, coincides with a separation of 10 mutual Hill radii for \textsc{Model J} and 100 mutual Hill radii for \textsc{Model E}, respectively \citep[see,][]{2017MNRAS.470.4337C}. The outermost planet therefore has a semi-major axis comparable to $40\%$ of Sedna's semi-major axis \citep{2016ApJ...824L..22M}. In the absence of external perturbations, our two models are stable \citep[as analytically and numerically verified in][]{2007ApJ...666..423Z} far beyond the total simulation time used in this study. In \textsc{Model J}, planet-planet scattering plays an important role in the dynamical evolution; in \textsc{Model E}, mutual planet-planet interactions are weak enough to be ignored. \textsc{Model E} serves as a comparison group to disentangle the contributions by external stellar flybys and internal planet-planet scatterings. 

We simulate the dynamical evolution of the stars and planets in two subsequent steps using the same approach in \cite{2017MNRAS.470.4337C}. In the first step, the equations of motion of the stars are integrated using \texttt{NBODY6++GPU} \citep{1999JCoAM.109..407S, 2003gnbs.book.....A, 2015MNRAS.450.4070W}. The positions, velocities, accelerations and the first derivative of accelerations are recorded using the Block Time Step Storage Scheme \citep{2010MNRAS.401.1898F,2012NewA...17..520F,2015ApJS..219...31C}, which allows us to reconstruct stellar flybys precisely down to the time resolutions of a few days (comparable to the integration time steps used by the planetary system integrator). In the second step, we use the \textsc{IAS15} integrator \citep{2015MNRAS.446.1424R} from the \texttt{rebound} package \citep{2012A&A...537A.128R} to integrate planetary systems. The precalculated perturbation data from step one is then communicated to the integrator using the \texttt{AMUSE} framework\footnote{\href{https://github.com/amusecode/amuse}{https://github.com/amusecode/amuse}} \citep{2009NewA...14..369P, 2013CoPhC.183..456P, 2013A&A...557A..84P}. This strict separation of stellar and planetary interactions enables us to separately study the effect of different planetary systems in the same star cluster and thereby disentangle the contributions from external stellar encounters and internal planet-planet scatterings. As a convenient side effect, it speeds up the calculations enormously, because we do not have to integrate the equations of motion of all the planets in the mutual gravitational field. Each planetary system is integrated for 50~Myr. This two-step integration scheme is embarrassingly parallel, carried out automatically using the \texttt{SiMon} automated job scheduling and monitoring toolkit \citep{2017PASP..129i4503Q}.
% * <spz@strw.leidenuniv.nl> 2017-09-14T06:25:58.268Z:
%
% ^.

\section{Results}
\label{sec:results}
\begin{figure}
	\centering
	\includegraphics[scale=0.45]{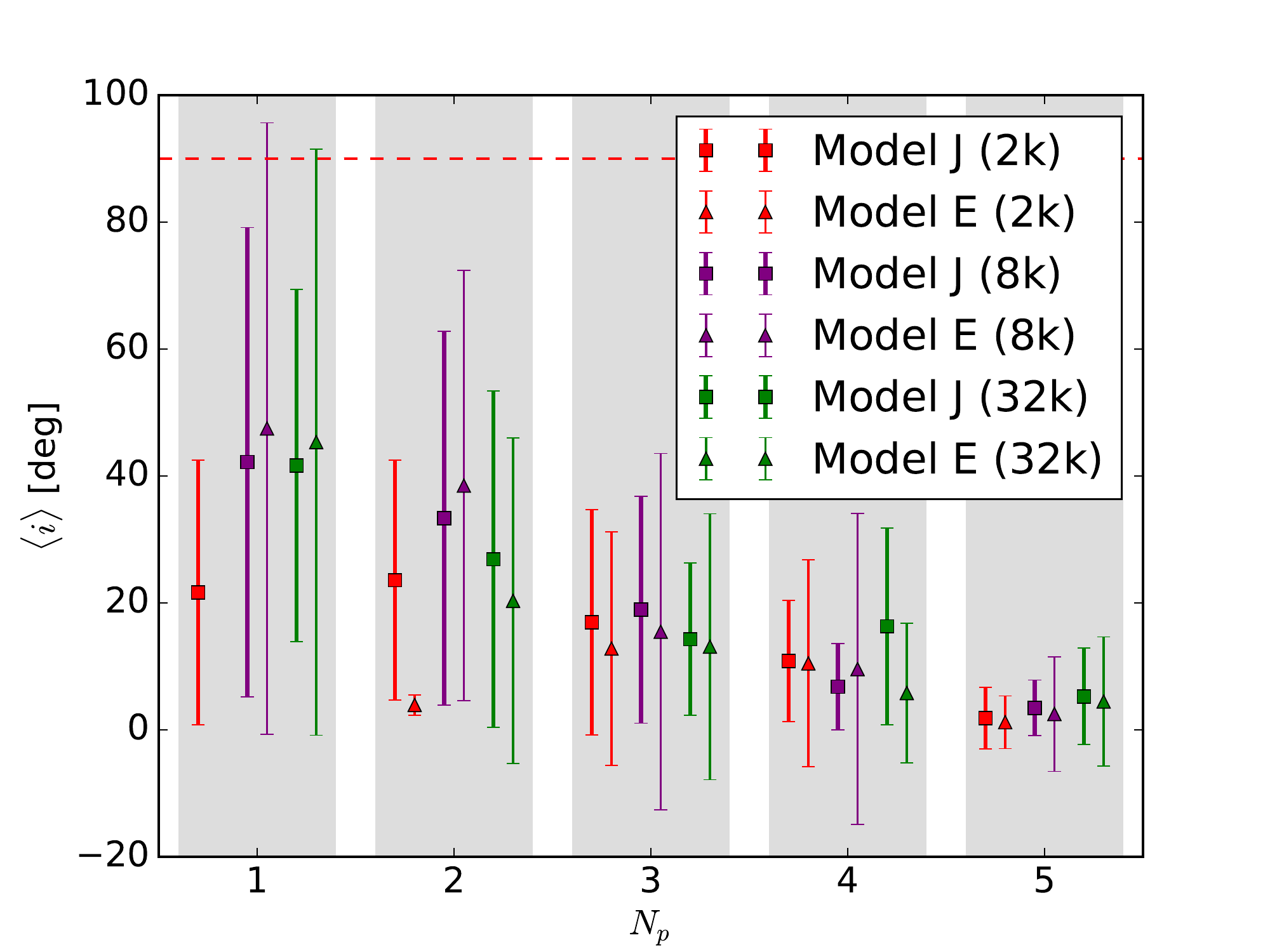}
	\caption{The mean inclinations $\langle i \rangle$ and their standard deviation $\sigma(i)$ (shown as errorbars) as a function of multiplicity $N_{\rm p}$ at $t=50$~Myr when the simulations finish. The dashed horizontal line divides the regimes of prograde orbits ($i < 90^{\circ}$) and retrograde orbits ($i > 90^{\circ}$). For the simulation data, the number of planetary systems having $N_{\rm p}$ planets ($N_{\rm p} = 1, 2, ..., 5$) is summarized in Table~\ref{tab:frac_np}. The values at $N_{\rm p}=5$ have the best statistical quality thanks to the adequate number of planetary systems in this bin. Due to the low stellar density in the $N=2{\rm k}$ cluster, the result in this cluster has little statistical significance for $N_{\rm p}<4$ systems. Note that the errorbars are showing the standard deviations rather than errors, and therefore the negative inclinations at the lower tips of the errorbars do not mean that some planets get negative inclinations.}
	\label{fig:inc_dispersion_np}
\end{figure}

% data table?
We carry out 200 \textsc{Model J} systems and 200 \textsc{Model E} systems in each host star cluster, and compare the statistical results. We aim to establish how the signatures from the parental cluster differ when planetary systems are exposed to different stellar densities, and to test whether planet-planet scattering can erase the signatures left by the parental clusters. Figure~\ref{fig:inc_dispersion_np} shows the correlation between mean inclinations (the statistical feature of mutual inclinations are given as $\sigma(i)$, i.e., the standard deviation of inclinations, shown as vertical bars) as a function of the number of planets after the planetary systems evolve in their parental cluster for $t=50$~Myr. 

Regardless of the density of the host cluster, both \textsc{Model J} and \textsc{Model E} exhibit the same statistics such that a higher degree of multiplicity results in a smaller mean inclination\footnote{Our definition of inclination is slightly different from the convention used in the exoplanet catalog: we define $i=0^{\circ}$ if the planet's orbital angular momentum normal is aligned with the spin angular momentum of the host star, which coincides with the edge-on exoplanets observed in transit surveys. In contrast, the inclination in the exoplanet catalog is $i=90^{\circ}$ in the edge-on case.} and in a smaller standard deviation (shown as errorbars). However, we are cautious to make direct comparison between the results of the $N=2{\rm k}$ cluster and $N=32{\rm k}$ cluster: due to the low stellar density in the $N=2{\rm k}$ cluster, most planetary systems are able to retain all their planets, and therefore very few poor planetary systems are produced (Table~\ref{tab:frac_np} summarizes the number of planetary systems as a function of $N_{\rm p}$ when the simulations end). Consequently, the $N=2{\rm k}$ result suffers from low-number statistics for the regime of $N_{\rm p} \leq 4$. In particular, the data point for the $N=2{\rm k}$ \textsc{Model E} system at $N_{\rm p}=2$ has no statistical significance because there is only one such planetary system. Fortunately, the statistical data at $N_{\rm p}=5$, as listed in Table~\ref{tab:np5_stat}, have the best quality thanks to the adequate samples. Based on these data, we observe that denser stellar environments indeed result in higher mean inclinations and larger standard deviation of inclinations. Nevertheless, it is interesting to notice that the difference is within the same order of magnitude, which indicates that the signature of inclination excitation is robust regardless of the stellar density in the parental cluster.

\begin{figure}
	\centering
	\includegraphics[scale=0.45]{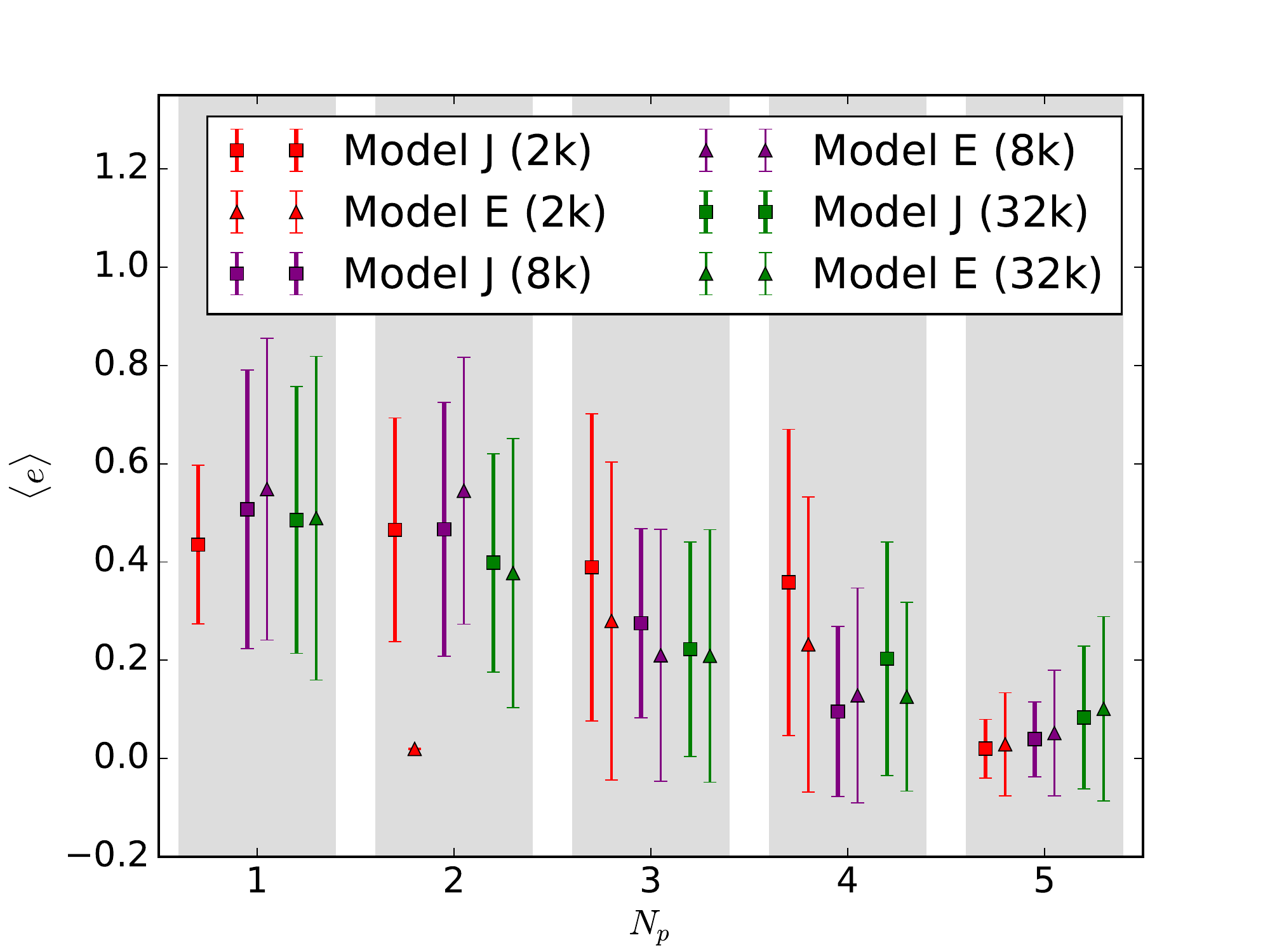}
	\caption{Same as Figure~\ref{fig:inc_dispersion_np}, but for the mean eccentricities $\langle e \rangle$ and their standard deviation $\sigma(e)$ as a function of number of planets $N_{\rm p}$. Note that the errorbars are showing the standard deviations rather than errors, and therefore the negative eccentricities at the lower tips of the errorbars do not mean that some planets get negative eccentricities.}
	\label{fig:ecc_dispersion_np}
\end{figure}

Considering that the planetary systems in our simulations start as circular and coplanar five-planet configurations, the eventual number of planets, $N_{\rm p}$, by the end of the simulation is an indicator of the degree by which the systems are perturbed, either by stellar fly-bys or by subsequent internal secular perturbations. Evidences of the intensive external perturbations are imprinted on the planetary systems in their high mean inclination and their mutual inclinations. Rich planetary systems, in contrast, remain relatively untouched compared to poor planetary systems. These differences are visible in the preserved nearly circular and coplanar orbits in the former population. This trend is visible in Figure~\ref{fig:ecc_dispersion_np} in which we present the mean eccentricities as a function of $N_{\rm p}$.

\begin{table}
\centering
	\begin{tabular}{c|c|c|c|c|c|c|c|c}
		\hline
		\hline
		\multirow{2}{*}{$N$} & \multirow{2}{*}{Model} & \multicolumn{7}{c}{$N_{\rm p}$} \\
		      & & {\bf 0} & {\bf 1} & {\bf 2} & {\bf 3} & {\bf 4} & {\bf 5} & {Total} \\
		\hline
		%$2{\rm k}$ & \textsc{J} &  0.01 & 0.03 &  0.045 & 0.04 & 0.02 & 0.855 \\
		$2{\rm k}$ & \textsc{J} &  2 & 6 &  9 & 8 & 4 & 171 & 200 \\
		\hline
		%$2{\rm k}$ & \textsc{E} &  0.01 &  0.00 & 0.005 & 0.015 & 0.035 & 0.935 \\
		$2{\rm k}$ & \textsc{E} &  2 &  0 & 1 & 3 & 7 & 187 & 200 \\
		\hline
		%$8{\rm k}$ & \textsc{J} &  0.01 & 0.03 &  0.045 & 0.04 & 0.02 & 0.855 \\
		$8{\rm k}$ & \textsc{J} &  7 & 17 & 27 & 14 & 20 & 115 & 200 \\
		\hline
		%$8{\rm k}$ & \textsc{E} &  0.01 &  0.00 & 0.005 & 0.015 & 0.035 & 0.935 \\
		$8{\rm k}$ & \textsc{E} &  3 &  8 & 7 & 11 & 31 & 140 & 200 \\
		\hline
		%$32{\rm k}$ & \textsc{J} &  0.095 &  0.185 & 0.245 & 0.10 & 0.11 & 0.265 \\
		$32{\rm k}$ & \textsc{J} &  19 &  37 & 49 & 20 & 22 & 53 & 200 \\
		\hline
		%$32{\rm k}$ & \textsc{E} &  0.035 &  0.04 & 0.085 & 0.145 & 0.265 & 0.43 \\
		$32{\rm k}$ & \textsc{E} &  7 &  8 & 17 & 29 & 53 & 86 & 200 \\
		\hline
		\hline
	\end{tabular}
    \caption{Number planetary systems with $N_{\rm p}$ planets at the end of the simulation ($T=50$~Myr). A planetary system with $N_{\rm p}=0$ means that all its planets are removed, and $N_{\rm p}=5$ means that all its planets are retained. The number of particles in the host cluster is denoted as $N$. Each ensemble has 200 initially identical planetary systems. A total of 1,200 planetary systems are integrated. }
    \label{tab:frac_np}
\end{table}

\begin{table}
\centering
	\begin{tabular}{c|c|c|c|c|c}
		\hline
		\hline
        $N$ & Model & $\langle e \rangle$ & $\sigma(e)$ & $\langle i \rangle$ & $\sigma(i)$ \\
        \hline
        2k & J & 0.020066 & 0.059425 & 1.853063 & 4.877858 \\
        \hline
        2k & E & 0.028947 & 0.104846 & 1.204652 & 4.157295 \\
        \hline
        8k & J & 0.039397 & 0.076209 & 3.461727 & 4.395462 \\
        \hline
        8k & E & 0.051723 & 0.127648 & 2.496239 & 9.037603 \\
        \hline
        32k & J & 0.083320 & 0.145129 & 5.274830 & 7.600131 \\
        \hline
        32k & E & 0.101066 & 0.187802 & 4.463308 & 10.182248 \\
        \hline
        \hline
	\end{tabular}
    \caption{The mean inclination $\langle i \rangle$ [deg], standard deviation of inclination $\sigma(i)$ [deg], mean eccentricity $\langle e \rangle$, and standard deviation of eccentricity $\sigma(e)$ of rich planetary systems in star clusters with $N_{\rm p}=5$. The number of planetary systems with $N_{\rm p}=5$ is listed in Table~\ref{tab:frac_np}.}
    \label{tab:np5_stat}
\end{table}

We use the term ``hot'' to describe planetary systems with high mutual inclinations and high eccentricities, and ``cool'' for those systems with relatively circular and coplanar orbits. In this context, all planetary systems start with a dynamically cool configuration, corresponding to a minimum angular momentum deficit (AMD, which is defined as the part of angular momentum resulted from non-circular and non-planar motion, see \citealt{1997A&A...317L..75L, 2000PhRvL..84.3240L, 2017A&A...605A..72L}). Large values of AMD usually lead to very chaotic behavior \citep{2011ApJ...735..109W}. In our simulations, the AMD of planetary systems is apparently injected by stellar encounters. Outer planets with large semi-major axes are more vulnerable to extended stellar perturbations \citep[e.g.,][]{2009ApJ...697..458S, 2017MNRAS.470.4337C}, and therefore they are the first ones to acquire an AMD. When the AMD is sufficiently large, orbit crossing will occur, and outer planets can share their AMD with the inner planets (given that planet-planet interactions are important), and the system has a tendency towards AMD equipartition \citep{2011ApJ...735..109W}. As such, when stellar encounters inject AMD into the outskirts of planetary systems, outer planetary systems in \textsc{Model J} need to share their AMD with inner planets, whereas in \textsc{Model E} the planetary systems can absorb all the AMD they receive from the parental star cluster. When a planet is ejected, it carries away both its own AMD contributed by stellar encounters, and also part of the AMD contributed by other planets. This explains the fact that the dispersion of inclinations is larger for \textsc{Model E} systems, as seen in Figure~\ref{fig:inc_dispersion_np}, and similarly observed in Figure~\ref{fig:ecc_dispersion_np} for the dispersion of eccentricities.

\begin{figure*}
	\centering
	\includegraphics[scale=0.6]{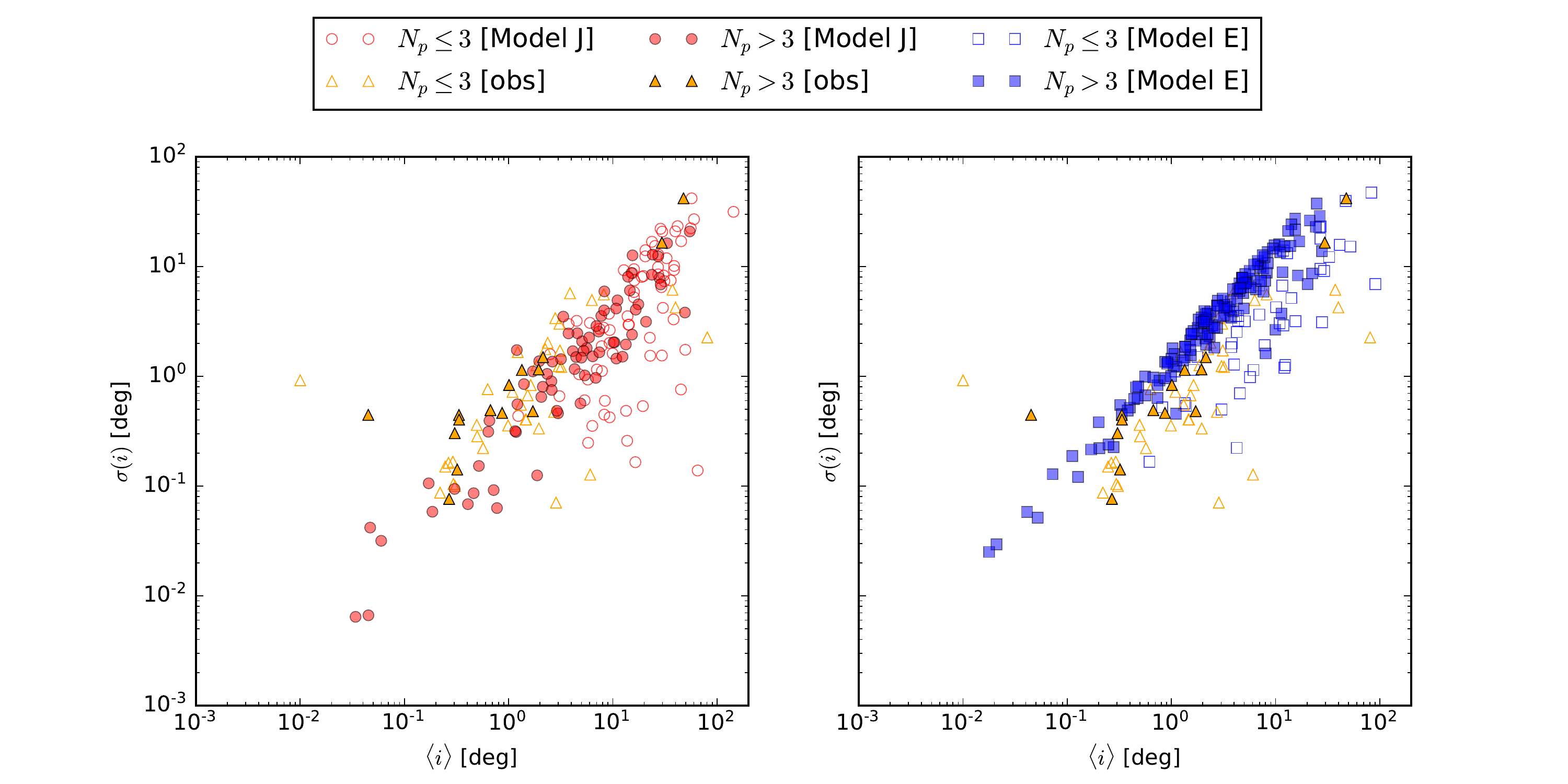}
	\caption{The standard deviation of inclinations per planetary system as a function of the mean eccentricity of that system (in the $N=32{\rm k}$ cluster). Left panel: comparison between \textsc{Model J} systems and observational data; Right panel: comparison between \textsc{Model E} systems and observational data. \textsc{Model J} systems are indicated with red markers; \textsc{Model E} systems are indicated with blue markers. Open markers indicate planetary systems with $N_{\rm p} \le 3$; filled markers indicate planet systems with $N_{\rm p} > 3$.  }
	\label{fig:inc_dispersion_per_sys}
\end{figure*}

In Figure~\ref{fig:inc_dispersion_per_sys}, we show the dispersion in the inclinations as a function of the mean inclination for \emph{each} planetary system in the $N=32{\rm k}$ cluster at $t=50$~Myr. 
The planetary systems in \textsc{Model E} with $N_{\rm p} > 3$ are are more concentrated towards the diagonal line, whereas systems with fewer planets tend to be more spread out. The rich systems show an approximately linear correlation between $\log(\langle i\rangle)$ and $\log(\sigma(i))$. These systems are only perturbed by external stellar encounters (i.e., planet-planet interactions are rare and have negligible effects on the orbits of the planets). The initial distribution of inclinations and eccentricities are well preserved for these systems, and even after 50\,Myr of dynamical evolution their initial conditions are still reflected in their orbital topologies. 
This is contrasted by the results of \textsc{Model J}, in which planets have masses comparable to Jupiter. In these models, external perturbations are enhanced by the internal secular evolution of the planetary systems.

\begin{figure*}
	\centering
	\includegraphics[scale=0.6]{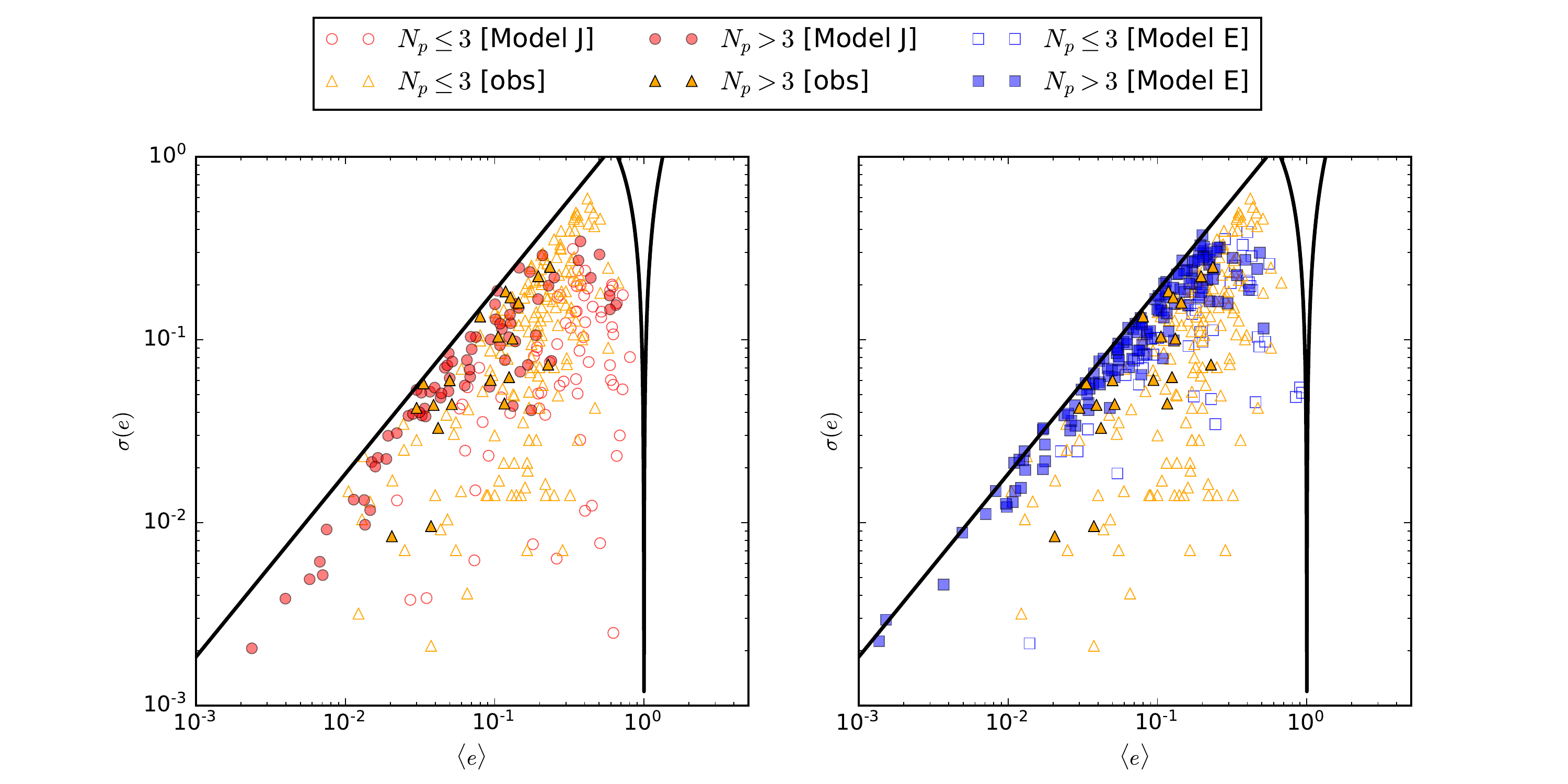}
	\caption{Same as Figure~\ref{fig:inc_dispersion_per_sys}, but for eccentricities. The diagonal thick black line coincides with the migration path of planetary systems that are solely driven by external stellar perturbations; the V-shape thick black curve is a schematic illustration of a possible migration pathway of planetary systems when solely driven by planet-planet scatterings, which in term have the feature of antiphase variation of eccentricity due to angular momentum conservation. }
	\label{fig:ecc_dispersion_per_sys}
\end{figure*}

In Figures~(\ref{fig:inc_dispersion_per_sys}--\ref{fig:ecc_dispersion_per_sys}), we see that these internal dynamical processes lead to a more diffuse distribution in inclinations and eccentricities for poor planetary systems ($N_{\rm p} \le 3$) compared those in \textsc{Model E}. In addition, rich planetary systems, in both models, also show a more confined distribution clustering near the diagonal line.

%These system have roughly the same mean inclinations $<i>$ as compared with top-layer systems, but their inclination dispersion $\delta i$ are lower because of much more efficient planet-planet scatterings (and therefore more efficient AMD sharing).

%to investigate how an individual planetary system (represented as a data point) migrate in the $<i>-\delta i$ space and the $<e>-\delta e$ space as functions of time. Recall that \textsc{Model J} and \textsc{Model E} both start with circular and coplanar configurations, therefore the mean inclination ($<i>$), mean eccentricity ($<e>$), dispersion of inclination ($\delta i$) and dispersion of eccentricity ($\delta e$) of any individual planetary system should be zero at $t=0$. 

\begin{figure*}
	\centering
	\includegraphics[scale=0.6]{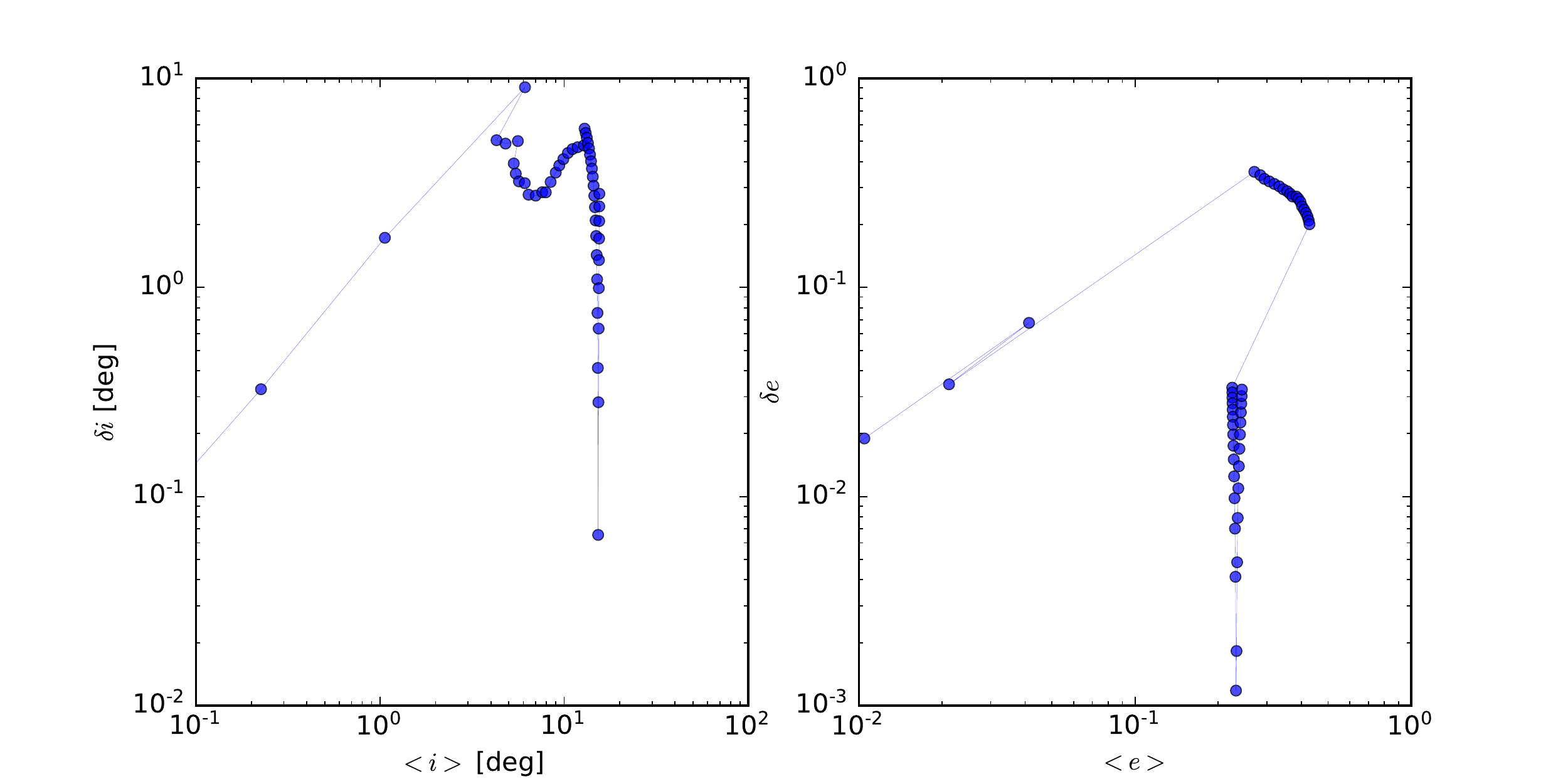}
	\caption{The migration pathways of inclination (left panel) and eccentricity (right panel) of a \textsc{Model E} system in the $N=32{\rm k}$ cluster. Each Myr checkpoint is denoted with a node in the evolution track. In this system, External stellar perturbations initially dominate the dynamical evolution of this system, causing the system to migrate alone the diagonal pathway (cf.~Fig.~\ref{fig:ecc_dispersion_per_sys}). Subsequent inclination excitations cause the system to deviate from the original diagonal path. Following the ejection of P2 and P3 at $t \sim 36$~Myr, the remaining two planets (P0 and P1) are weakly coupled. The conservation of angular momentum causes the system to migrate alone a V-shape trajectory.}
	\label{fig:inc_dispersion_migration}
\end{figure*}

The broad distribution is formed through the combined effects of external perturbations and internal planetary system dynamics. The inclinations of the orbits of the planets are affected by close encounters with other stars. In the same planetary system, the outermost orbit is most strongly affected by stellar encounters. Therefore, during the onset of an encounter, the orbital inclinations of all planets change systematically, and outer planets experience more significant changes than inner planets, leading a systematic growth of both the mean inclination and the standard deviation of inclinations. Planetary systems that experience external perturbations therefore move along the black diagonal line in Figure~\ref{fig:ecc_dispersion_per_sys} so long as internal planetary dynamics can be neglected. On the other hand, if planet-planet scattering is the only driver of the variation in eccentricity, then due to the conservation of angular momentum, the evolution of the eccentricity exhibits an anti-phase sinusoidal pattern in which case the planetary system migrates alone a V-shape trajectory (denoted with the black curve in Fig.\,~\ref{fig:ecc_dispersion_per_sys}). In a real environment, external perturbations and internal dynamics interplay when the planetary system co-evolves with its parental cluster. Stronger internal dynamical scatterings lead to a broader distribution in $\langle e\rangle$ and $\sigma(e)$. This can be observed in \ref{fig:ecc_dispersion_per_sys} in  \textsc{Model J} in which planetary systems experience stronger internal interactions compared to those in \textsc{Model E}. As a consequence, the spread of orbital elements is more diffuse, as can be seen the same figure. Among systems in the same model, the poor planetary systems with $N_{\rm p} \le 3$ are more strongly perturbed by external encounters compared to the rich planetary systems ($N_{\rm p} > 3$), which leads to enhanced frequencies of orbit-crossings and ultimately stronger planet-planet scattering. As observed in Figure~\ref{fig:ecc_dispersion_per_sys}, the poor population tends to exhibit a broader distribution in $\langle e\rangle$ - $\sigma(e)$ than the rich systems. We illustrate the evolution of a typical planetary system that got perturbed by a passing star in Figure~\ref{fig:inc_dispersion_migration}.
Figure~\ref{fig:ecc_dispersion_np} highlights the evolution of orbital elements driven by internal as well as by external perturbations of the same planetary system shown in Figure~\ref{fig:inc_dispersion_migration}. In addition, the system undergoes rapid changes in $\langle i\rangle$ and $\sigma(i)$ when two planets are ejected between 30-40~Myr.

We conclude that external perturbations lead to the excitations of mean inclination and mean eccentricity, while internal dynamical evolution leads to variations in the dispersions of the inclination and the eccentricity. This is also shown in Figures~\ref{fig:inc_dispersion_per_sys} and \ref{fig:ecc_dispersion_per_sys}, where we find that planetary systems spread beneath the diagonal line, implying that the parental clusters have left an imprint in the early stage of the evolution of its planetary systems.

\begin{figure*}
	\centering
	\includegraphics[scale=0.7]{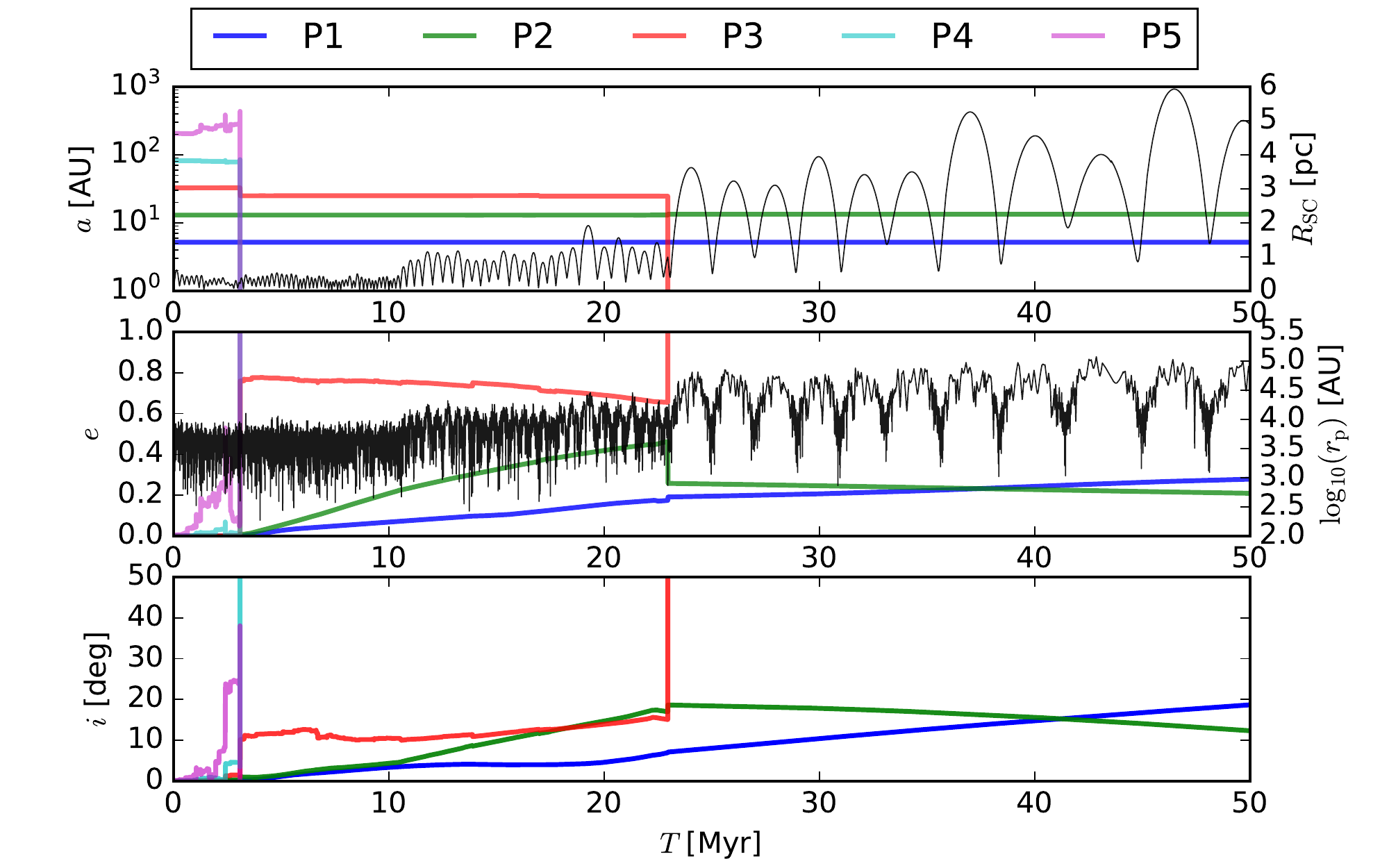}
	\caption{Same planetary system in Figure~\ref{fig:inc_dispersion_migration}, but for the evolution of $a$, $e$, and $i$ as a function of time. Individual planets are plotted with colored curves. The thin gray curve at the top panel depicts the distance from the cluster center to the planetary system (y-axis on the right); the think gray curve on the middle panel depicts the distance from the closest perturber to the planetary system host star (y-axis on the right). From $t \sim 0-10$~Myr, the planetary system locates near the cluster center, and the strong and frequent perturbations in this region cause efficient excitation and ejection of two outermost planets (P4 and P5) at $t \sim 3$~Myr. A gravitational slingshot at $t \sim 23$~Myr causes the ejection of another planet (P3), and throws the planetary system into an eccentric cluster orbit. Consequently, the planetary system periodically dives into the cluster center, where the perturbation frequency and strength both enhanced. The two remaining innermost planets (P1 and P2) are weakly coupled; both eccentricities and inclinations exhibit antiphase variation due to the conservation of angular momentum, which in turn produce the V-shape in Figure~\ref{fig:inc_dispersion_migration}. Interestingly, if an observer were to observe the system at $t \sim 40$~Myr, they will find the two planets coplanar; if observed at around $t \sim 25$~Myr, the two planets have a mutual inclination of $\sim 10^{\circ}$.}
	\label{fig:ecc_acc_semi_rsc}
\end{figure*}

\section{Discussion}
\label{sec:discussion}
% natural selection process?
% the disk can be initially inclined and eccentric
% observational biases
% star cluster properties, p sys properties, time spent in the cluster

The Solar System, as well as rich planetary systems such a TRAPPIST-1, are systematically flat and circular. The mutual inclinations of these multi-planet systems are less than $5^{\circ}$, and the inclination distribution can be described with a Rayleigh distribution \citep{2012AJ....143...94T,2012ApJ...761...92F, 2016ApJ...816...66B}. In terms of eccentricities, \cite{2015ApJ...808..126V} find that the eccentricity in \emph{Kepler} multi-planet systems are low and can also be described by a Rayleigh distribution with $\sigma = 0.049 \pm 0.013$. \cite{2016PNAS..11311431X} point out that there is an eccentricity dichotomy among \emph{Kepler} planets: the mean eccentricity of single-transit planets is $\langle e \rangle \sim 0.3$, but $\langle e \rangle$ drops drastically for multi-transit systems. On the other hand, \cite{2010ARA&A..48...47A}, for example, suggests that the Solar System is formed in the outskirts of a star cluster with a few thousand member stars.  This is in agreement with the results in \cite{2017MNRAS.470.4337C}, where they find that planetary systems in the outskirts of star clusters are relatively unperturbed because of the low stellar density there, but inconsistent with the possible abduction of the dwarf planet Sedna from another star in the Sun's birth cluster \citep{2015MNRAS.453.3157J}.
We therefore suspect that the Kepler dichotomy implies a natural selection process: those systems with high degrees of multiplicity will automatically be dynamically cooler because they did not undergo considerable external perturbations; those systems with few planets are naturally hotter, because they went through strong external perturbations, and the observed planets in these systems (most likely short-period planets) are the survivors of this process. 

% spin-orbit misalignment: Kozai, planet-planet interaction, mass accreation
In the Solar System, while most planets are roughly on the same orbital plane, the reason for the misalignment of the Ecliptic plane with respect to the Sun's equatorial plane by $7^{\circ}$ still unclear. Similar systems such as Kepler-56 are observed, in which two planets are nearly coplanar but have large misalignment ($>37^{\circ}$) with the spin of the host star \citep{2013Sci...342..331H}. Various mechanisms have been proposed to explain the formation of these systems, for example, through an inclined, non-transiting companion \citep[e.g., ][]{2014ApJ...794..131L, 2016AJ....152..165O, 2017MNRAS.468.3000M}, or through an inclined protoplanetary disk \citep[e.g.,][]{2017A&A...604A..88W}. Another quite distinct systems shows a typical similarity with Kepler-108 \citep[e.g.,][]{2017AJ....153...45M}, shows high degree of mutual inclinations. 
However, according to \cite{2015MNRAS.453.3157J} the misalignment can be explained by a steady flow of gas when the Solar System moved through an ambient gaseous medium in its infancy.
In fact, these two distinct systems can be produced in our simulations through repeated hyperbolic and nearly-parabolic encounters in star clusters. The system shown in Figure~\ref{fig:inc_dispersion_migration} can be either Kepler-56 analogue or Kepler-108 analogue, depending on the time of the observation: when observed at $t \sim 30$~Myr and $t \sim 50$~Myr, the two planets have high mutual inclination, resembling Kepler-108; when observed at $t \sim 40$~Myr, the two planets in the system are nearly coplanar, and they have considerable misalignment with respect to the spin of the host star, which resembles Kepler-56.

% transit probability
% The probability that an exoplanet transits its host star drops off rapidly with the increment of semi-major axis \citep{2010ApJ...712.1433B}.

\section{Conclusions}
\label{sec:conclusions}
Most star and planets are formed in star clusters. Planetary systems in the field, including our Solar System, may have spent their early times in the parental clusters. Consequently, newly formed planetary systems in star clusters are subject to frequent external perturbations due to the high stellar density, which in turn leave signatures in the orbital elements. In this study, we investigate such signatures using direct $N$-body simulations. Based on our simulation results, we argue that the ``Kepler dichotomy'' may be such a signature imprinted by the parental cluster. Our main conclusions are:

\begin{itemize}
\item The birth environments of a planetary system in the field can be constrained by the observed orbital elements of its planets. Star cluster environments impose a natural selection process to the planetary systems formed within. While most presently known planets are detected in the field, they are essentially survivors of the stellar encounters in their parental clusters. 

\item In our simulations, the mean inclination $\langle i\rangle$ and mean eccentricity $\langle e\rangle$ of a planetary system decline as a function of multiplicity $N_{\rm p}$. Rich planetary systems ($N_{\rm p} > 3$) are mostly formed outside the half-mass radii their parental clusters, as the low stellar densities in those regions leave them mostly unperturbed. In contrast, poor planetary systems ($N_{\rm p} \le 3$) are mostly formed in the high-density central regions of the parental cluster. As survivors of intensive perturbations, these systems are usually compact and with moderate-to-high orbital eccentricities and mutual inclinations. 

\item We simulate multi-planet systems in different stellar densities, and found that dense stellar environments produce slightly stronger signatures of mean inclination/eccentricity excitation among rich planetary systems. However, the anticorrelation between $N_{\rm p}$ and $\langle i\rangle$ (and likewise between $N_{\rm p}$ and $\langle e\rangle$) is robust regardless of the density of the stellar environments.

\item The total angular momentum deficit (AMD) of a planetary system is another signature imprinted by stellar encounters in the parental clusters. Modern planet formation theories suggest that planets form in circular and coplanar orbits, corresponding to the maximum angular momentum. The total AMD of a system can increase when stellar encounters excite $e$ and $i$, but will remain mostly unchanged for planetary systems that are not externally perturbed. After a planetary system escapes the parental cluster, its AMD remains static.

\item When external perturbations dominate the dynamical evolution of a planetary system, the mean inclination $\langle i\rangle$ and the dispersion in the inclination $\sigma(i)$ grows systematically (likewise for the mean eccentricity $\langle e\rangle$ and the dispersion of eccentricity $\sigma(e)$). When internal evolution dominates the dynamical evolution, $i$ and $e$ of individual planet orbits exhibit an anti-phase oscillation due to the conservation of angular momentum. The former mechanism causes the system to migrate along a diagonal line in $\langle i\rangle$--$\sigma(i)$ space and in  $\langle e\rangle$--$\sigma(e)$ space. This exchange of angular moment among planets, cause the system to migrate a V-shape trajectory in these spaces. As such, the evolution in $\langle i\rangle$ and $\sigma(i)$, but also in $\langle e\rangle$ and $\sigma(e)$ can be used to estimate the contributions of external perturbation with respect to the internal secular dynamical evolution of individual planetary systems.

\item Planetary systems with high mutual inclinations, such as Kepler-108, and for which the orbits are co-planar and inclined, such as Kepler-56 and the Solar System, can be reproduced in our simulations. We speculate that these two classes of systems may actually belong to the same family, but observed at a different epoch.
\end{itemize}

Due to the restriction in computational resources, the build-up of errors and the chaotic nature of $N$-body systems, our study is limited to planetary systems with relatively wide orbits; the inner most planet is at 5.2\,AU (the outermost planet is at $a \sim 200$~AU. This setting makes planetary systems sensitive to external perturbations, and therefore allows us to produce the signatures of the parental clusters without having to carry out simulations for extensive periods of time. Our investigation focused on clusters in which the primordial gas has already depleted, and we ignored tidal effects between planets and their stellar host. The actual distribution of mean inclinations/eccentricities as a function of $N_{\rm p}$ may be smaller than our models, especially among tight planetary systems \citep[e.g., Kepler-11, see][]{2011Natur.470...53L} and/or in the scenarios where eccentricity damping by tidal effect and gas drag operate. Nevertheless, our results highlight the natural selection process in the parental cluster and its implication to observable quantities, which is important for understanding the planet formation and evolution processes.

\section*{Acknowledgements}

We thank the anonymous referee for insightful comments and suggestions. We thank Vincent van Eylen and Subo Dong for useful discussions. This work was supported by the Netherlands Research Council NWO (grant \#621.016.701 [LGM2]) by the Netherlands Research School for Astronomy (NOVA). This research was supported by the Interuniversity Attraction Poles Programme (initiated by the Belgian Science Policy Office, IAP P7/08 CHARM) and by the European Union's Horizon 2020 research and innovation programme under grant agreement No 671564 (COMPAT project).

%%%%%%%%%%%%%%%%%%%%%%%%%%%%%%%%%%%%%%%%%%%%%%%%%%

%%%%%%%%%%%%%%%%%%%% REFERENCES %%%%%%%%%%%%%%%%%%

% The best way to enter references is to use BibTeX:

\bibliographystyle{mnras}
\bibliography{bibtex} % if your bibtex file is called example.bib

% Alternatively you could enter them by hand, like this:
% This method is tedious and prone to error if you have lots of references
%\begin{thebibliography}{99}
%\bibitem[\protect\citeauthoryear{Author}{2012}]{Author2012}
%Author A.~N., 2013, Journal of Improbable Astronomy, 1, 1
%\bibitem[\protect\citeauthoryear{Others}{2013}]{Others2013}
%Others S., 2012, Journal of Interesting Stuff, 17, 198
%\end{thebibliography}

%%%%%%%%%%%%%%%%%%%%%%%%%%%%%%%%%%%%%%%%%%%%%%%%%%

%%%%%%%%%%%%%%%%% APPENDICES %%%%%%%%%%%%%%%%%%%%%

%%%%%%%%%%%%%%%%%%%%%%%%%%%%%%%%%%%%%%%%%%%%%%%%%%

% Don't change these lines
\bsp	% typesetting comment
\label{lastpage}
\end{document}